\documentclass{article}
\begin{document}
\title{The magnetic dipole interaction in Einstein-Maxwell theory}
\author{W.B.Bonnor}
%\address{Queen Mary, University of London\\London E1 4NS}
\maketitle

\begin{abstract}
I derive an exact,
static, axially symmetric solution of the Einstein-Maxwell equations
representing two massless magnetic dipoles, and compare it with the corresponding solution
of Einstein's equations for two massless spinning particles (see gr-qc/0201094).
I then obtain an exact stationary solution of the Einstein-Maxwell equations representing
two massless spinning magnets in balance.  The conclusion is that the spin-
spin force is analogous to the force between two magnetic dipoles, but of opposite sign, and that
the latter agrees with the classical value in the first approximation.
\end{abstract}

\begin{flushleft}
Comments: 5 pages; Journal reference: Class. Quantum Grav. 19 (2002) 149.
\end{flushleft}

\section{Introduction}
The resemblance of spin-spin interaction in general relativity and the interaction
between magnetic dipoles in Maxwell's theory was pointed out by Wald [1].
Wald based his theory of the spin-spin interaction on the equations
of motion for spinning test particles, and took for the magnetic
interaction expressions well known in classical electromagnetism.
In gr-qc/0201094, I gave an exact solution of Einstein's equations
exhibiting the spin-spin interaction; here I obtain the
corresponding static solution of Einstein-Maxwell theory for two magnetic dipoles
and exhibit the similarity between the two solutions.  Secondly, I use another solution
of the Einstein-Maxwell equations to show how two massless
spinning magnetic dipoles can balance.  The spacetime is throughout
assumed to be axially symmetric, i.e. the axes of
two magnets are collinear and may be parallel or antiparallel.

In section 2 I derive the exact static solution for two magnetic dipoles, and
in section 3 I write down the exact solution for two balancing spinning magnetic
dipoles.  Section 4 discusses the physical interpretation of the solutions,
and there is a concluding section 5.

\section{Exact solution for two magnetic dipoles}
The field equations are those of Einstein-Maxwell theory:
\begin{eqnarray}
R^{i}_{k}&=&2F^{ia}F_{ka}-\frac{1}{2}\delta^{i}_{k}F^{ab}F_{ab},\\
F_{ik}&=&A_{i,k}-A_{k,i},\\
F^{ik}_{;k}&=&0,
\end{eqnarray}
where $R^{i}_{k}$ is the Ricci tensor, $F_{ik}$ the electromagnetic field
tensor, and $A_{i}$ the vector potential; a comma denotes partial differentiation
and a semi-colon covariant differentiation.

The metric may be taken in Weyl static form
\begin{equation}
ds^{2}=-F^{-1}[e^{\mu}(dz^{2} +dr^{2})+r^{2}d\theta^{2}]+F dt^{2},
\end{equation}
where $F$ and $\mu$ are functions of $z$ and $r$.  The coordinates will be
numbered
\[ x^{1}=z, x^{2}=r, x^{3}=\theta, x^{4}=t,  \]
where
\[ \infty>z>-\infty, r>0, 2\pi\geq\theta\geq 0, \infty>t>-\infty,\]
$\theta=0$ and $\theta=2\pi$ being identified.  The vector potential takes
the form
\begin{equation}
A_{i}=(0,0,\psi,0)
\end{equation}
where $\psi$ is a function of $z$ and $r$.

The field equations (1)-(3) may be written in the form
\begin{eqnarray}
R_{11}+R_{22}&=&\mu_{11}+\mu_{22}-F^{-1}\nabla^{2}F+\frac{3}{2}F^{-2}(F_{1}^{2}+F_{2}^{2})=0,\\
R_{11}-R_{22}&=&r^{-1}\mu_{2}+\frac{1}{2}F^{-2}(F_{1}^{2}-F_{2}^{2})=2r^{-2}F(\psi_{2}^{2}-\psi_{1}^{2}),\\
2R_{12}&=&-r^{-1}\mu_{1}+F^{-2}F_{1}F_{2}=-4r^{-2}F\psi_{1}\psi_{2},\\
R_{4}^{4}-R_{3}^{3}&=&e^{-\mu}[-\nabla^{2}F+F^{-1}(F_{1}^{2}+F_{2}^{2})]=-2r^{-2}e^{-\mu}F^{2}(\psi_{1}^{2}+\psi_{2}^{2}),\\
\nabla^{*2}\psi&=&-F^{-1}(F_{1}\psi_{1}+F_{2}\psi_{2}),
\end{eqnarray}
where suffices 1 and 2 on the right mean differentiation with respect to $z$
and $r$ respectively, and
\begin{eqnarray*}
\nabla^{2}X&=&X_{11}+X_{22}+r^{-1}X_{2},\\
\nabla^{*2}X&=&X_{11}+X_{22}-r^{-1}X_{2}.
\end{eqnarray*}

Eqns (6)-(10) bear a close resemblance to (2)-(6) of [2], and from any
solution of the latter set one may generate a solution of the former set by
substituting
\begin{equation}
f\rightarrow F^{1/2},\;\;\;\;w\rightarrow i\psi,\;\;\;\;\;\nu\rightarrow \frac{1}{4}\mu,
\end{equation}
as one may verify by direct calculation.  \footnote{This generation procedure
was first used by me in [3], though I did not describe it explicitly.
An account of it was given by Kinnersley [4], and
it was generalised by Fisher in [5].}

Applying this transformation to the Papapetrou solution (8)-(12) of [2]
we generate a magnetic spacetime with the following metric coefficients:
\begin{eqnarray}
F^{-1}&=&\cosh^{2}\xi_{1},\\
i \psi&=&r\xi_{2},\\
\mu_{1}&=&4r\xi_{11}\xi_{12},\\
\mu_{2}&=&2r[(\xi_{12})^{2}-(\xi_{11})^{2}],\\
\nabla^{2}\xi&=&0.
\end{eqnarray}
To get a real solution of our problem we choose an imaginary solution of (16):
\begin{equation}
\xi=-i\left(\frac{M_{1}}{R_{1}}+\frac{M_{2}}{R_{2}}\right),
\end{equation}
where $M_{1}$ and $M_{2}$ are moments of magnets placed on the $z$-axis at
$z=\pm b, b>0$ and pointing in the + or - $z$-direction, and $R_{1}=\mid [(z-b)^{2}+r^{2}]^{1/2} \mid,
\newline R_{2}=\mid [(z+b)^{2}+r^{2}]^{1/2} \mid $.

Substituting (17) into (12)-(15) we obtain
\begin{eqnarray}
F&=&\sec^{2}Y,\\
\psi&=&r^{2}\left(\frac{M_{1}}{R_{1}^{3}}+\frac{M_{2}}{R_{2}^{3}}\right),\\
\mu&=&-\sum_{i=1}^{2}\frac{M_{i}^{2}r^{2}(9r^{2}-8R_{i}^{2})}{2R_{i}^{8}}
        +\frac{M_{1}M_{2}[3(r^{2}+z^{2}-b^{2})^{3}+2b^{2}r^{2}(9r^{2}+9z^{2}-b^{2})]}{2b^{4}R_{1}^{3}R_{2}^{3}}\nonumber\\
        +C,
\end{eqnarray}
where $C$ is an arbitrary constant, and where $Y$ is given by
\begin{equation}
Y=\frac{M_{1}(z-b)}{R_{1}^{3}}+\frac{M_{2}(z+b)}{R_{2}^{3}}.
\end{equation}
The physical meaning of this solution will be discussed in section 4, but I
note here its resemblance to the solution for spinning particles, (15)-(17)
of [2].

\section{Exact solution for spinning magnetic dipoles in balance}
This solution can be obtained from the Perjes-Israel-Walker (PIW) class [6] [7],
which describes the stationary field of sources bearing mass, charge,
magnetic moment and angular momentum.  The field equations are (1)-(3).  The PIW solutions are free from conical
singularities, but, in general they contain another singularity in the
neighbourhood of the rotation axis which I have recently [8] called a
torsion singularity.

A PIW solution for two particles was examined in [9], and I use it here
(with a slightly different notation).  To rid the solution of electric
charge (which is not under consideration in this paper) one must put
the parameters $m_{1},m_{2}$ equal to zero; this has the effect of abolishing
the masses of the particles and the torsion singularity.  The result is
\begin{eqnarray}
ds^{2}&=&-f^{-1}(dz^{2}+dr^{2}+r^{2}d\theta^{2})+f(wd\theta-dt)^{2},\\
\phi&=&-\epsilon f,\;\;\Psi=\epsilon Yf,
\end{eqnarray}
where
\begin{eqnarray}
f^{-1}&=&1+Y^{2},\\
w&=&2r^{2}\left(\frac{M_{1}}{R_{1}^{3}}+\frac{M_{2}}{R_{2}^{3}}\right),\\
\epsilon&=&\pm 1,
\end{eqnarray}
where $Y$ is as in (21).
Here $\phi$ is the electric potential $A_{4}$, but $\Psi$ is the magnetic
{\em scalar} potential.  This solution will be explained in the next section.

\section{Physical interpretation}
In this section I discuss the exact solutions in the previous two sections,
starting with the one in section 2.

We first notice that although $F$ tends to unity and the metric becomes asymptotically flat as $R_{1}$ and $R_{2}$ tend to
infinity, $F$ becomes infinite when
\begin{equation}
Y:= \frac{M_{1}(z-b)}{R_{1}^{3}}+\frac{M_{2}(z+b)}{R_{2}^{3}}=\pm \frac{\pi}{2},
\end{equation}
and one must regard this as the locus of a singularity surrounding the
magnets.  Our main interest in this paper is in the portion of the $z$-axis
$-b<z<b$ where a conical singularity will enable us to deduce the
force of interaction between the magnets.  It is easy to see that if
$M_{i}/b^{3}, i=1,2$ are sufficiently small there is always a part of the $z$-axis between the particles
which is outside the singular surface (27), and the latter bifurcates into two
surfaces, one around each magnet.  I shall henceforth suppose this condition
satisfied, and confine attention to that part of $-b<z<b, r=0$ which is
outside (27).

Using (4) and (18) we have
\[ g_{44}=1+Y^{2}+O(R^{-8}), \]
where $R^{2}=z^{2}+r^{2}.$   This contains no term of order $R^{-1}$, so
there is no term representing mass: the magnets must be considered massless.
>From (19) we see that the magnetic vector potential has at infinity the
form expected for two magnets, parallel or anti-parallel, at $z=\pm b, r=0.$
The expression for $\mu$ can be dealt with as was $\nu$ in [2].  It
represents a conical singularity either between the magnets, or outside them,
according to the choice of $C$.  Choosing the former alternative we can, as
before, regard the conical singularity as a cosmic string and calculate the
force which it represents.  It turns out to be
\begin{equation}
p=\frac{3M_{1}M_{2}}{8b^{4}},
\end{equation}
plus higher terms of order $M_{1}^{2}M_{2}^{2}/b^{8}$.
Comparing this with (23) of [2] we see that the force between magnets is
similar to the force between spinning particles, but opposite in sign.
This agrees with Wald [1].  It is the same as that found in elementary
magnetism [10].  This, incidentally, gives credibility to the
cosmic string calculation of the force, which was put forward tentatively
in [2].

This concludes the interpretation of the exact solution in section 3, and I
turn now to the one in section 4.  From the metric (22), in particular from
the $d\theta dt$ term, we see that the solution refers to two
particles spinning with angular momenta $M_{1}, M_{2}$ on the axis of
symmetry at $z=\pm b$,  and from (21),(23) and (24) it follows that the particles carry
magnetic dipoles $M_{1}, M_{2}$, i.e. numerically the same as the angular momenta in relativistic units.  The absence of conical singularities shows that
the spin-spin and the magnetic forces balance, as expected from the
above discussion of the solution in section 2.  Once again the particles are
massless, as one sees by examining $g_{44}$ at infinity.

\section{Conclusion}
I have argued that the force between two collinear parallel or anti-parallel
magnets in Einstein-Maxwell theory is given approximately by (28), which is the same as the
corresponding form in classical magnetism.  Of course, in the E-M case
some ambiguity attaches to $b$ because of the conical singularity between
the particles.

The magnetic force has the same form as the spin-spin force found in [2],
but opposite sign.  This is compatible with the interpretation of the
solution in section 3, in which the magnetic moments and the angular
momenta are equal (in relativistic units).  It also agrees with the
more general approximate result of [8] in which it was shown that, in the
second approximation, the forces balance if $h_{1}h_{2}=M_{1}M_{2}$,
$h_{1},h_{2}$ being the angular momenta.

In this paper and also in [2] I have had to treat the particles as massless
because appropriate exact solutions with massive particles are unknown.
This means that, although spin-spin and magnetic dipole forces can be
treated by analysing conical singularities, the torsion singularity [8], which
arises through the coupling of mass and angular momentum, does not appear.

\section*{References}
{[1]} Wald R 1972 {\em Phys. Rev.} D{\bf 6} 406\\
{[2]} Bonnor W B 2002 {\em Class. Quantum Grav.} (submitted)\\
{[3]} Bonnor W B 1966 {\em Z. f\"{u}r Physik} {\bf 190} 444\\
{[4]} Kinnersley W 1974 in {\em General Relativity and Gravitation} ed. Shaviv G
and Rosen J (Tel Aviv)\\
{[5]} Fisher E 1979 {\em J. Math. Phys.} {\bf 20} 2547\\
{[6]} Perjes Z 1971 {\em Phys. Rev. Lett.} {\bf 27} 1668\\
{[7]} Israel W and Wilson G A 1972 {\em J. Math. Phys.} {\bf 13} 865\\
{[8]} Bonnor W B 2001 {\em Class. Quantum Grav.} {\bf 18} 2853\\
{[9]} Bonnor W B and Ward J P 1973 {\em Comm. Math. Physics} {\bf 28} 323\\
{[10]}Ferraro V C A 1962 {\em Electromagnetic Theory} (London: Athlone Press) page 242
\end{document}